\documentclass[twocolumn,showpacs,preprintnumbers,amsmath,amssymb,superscriptaddress,groupedaddress,prb]{revtex4-1}

\usepackage{epsfig}                                                            
\usepackage{graphicx}
\usepackage{dcolumn}
\usepackage{color}
\usepackage{bm}
\usepackage{subfigure} 
\usepackage{epsfig} 

\begin{document} 

\title{\bf Evolution of orbital phases with particle size in nanoscale stoichiometric LaMnO$_3$}

\author{Parthasarathi Mondal} \affiliation{Nanostructured Materials Division, Central Glass and Ceramic Research Institute, CSIR, Kolkata 700032, India}
\author{Dipten Bhattacharya} \email{dipten@cgcri.res.in}\affiliation{Nanostructured Materials Division, Central Glass and Ceramic Research Institute, CSIR, Kolkata 700032, India}
\author{Anwesha Maity} \affiliation{Non-Oxide Ceramics and Composite Division, Central Glass and Ceramic Research Institute, CSIR, Kolkata 700032, India}
\author{Omprakash Chakrabarti} \affiliation{Non-Oxide Ceramics and Composite Division, Central Glass and Ceramic Research Institute, CSIR, Kolkata 700032, India}
\author{A.K.M. Maidul Islam}  \affiliation{Surface Physics Division, Saha Institute of Nuclear Physics, Kolkata 700064, India}
\author{Manabendra Mukherjee} \affiliation{Surface Physics Division, Saha Institute of Nuclear Physics, Kolkata 700064, India}

\date{\today}

\begin{abstract}

The thermodynamically stable long-range orbital order in bulk LaMnO$_3$ becomes metastable at nanoscale around a critical particle size d$_C$ $\sim$20 nm. The orbital order-disorder transition switches from $\textit{reversible}$ to $\textit{irreversible}$ at d$_C$ while the resistance in the orbital ordered state decays by 2-4\% over a time scale of $\sim$3000s. At well below d$_C$, of course, a stable orbital disordered phase emerges. The orthorhombic distortion of the underlying crystallographic structure (space group Pbnm) decreases systematically with the decrease in particle size and at far below d$_C$ (e.g., at $\sim$10 nm), the structure becomes cubic (space group Pm$\bar{3}$m). Using the crystallographic and electrical resistance data, a phase diagram has been constructed showing the evolution of different orbital phases as a function of particle size across $\sim$10 nm to bulk for stoichiometric LaMnO$_3$.

\end{abstract}
\maketitle 

\section{Introduction} 
The issue of phase stability and transition in confined geometry (e.g., in nanoscale system) assumes importance both for its physics as well as in the context of design of different nanosized architectures for nanoelectronics. Although, this has been addressed, theoretically, nearly 35 years back \cite{Buffat}, direct measurement of the phase stability and transition enthalpy together with kinetics is being reported \cite{Dippel}$^-$\cite{Schmidt} only in recent times. Several novel approaches - such as scanning tunneling microscopy in association with perturbed angular correlation \cite{Dippel}, laser irradiation coupled with local calorimetry \cite{Zhang}, ultrasensitive nanocalorimetry \cite{Schmidt} etc - have been adopted for measuring the melting point and latent heat of melting in nanoscale atomic clusters. Deeper understanding of the phase transition thermodynamics and kinetics for electronic phase superstructures\cite{Tokura} - such as charge and orbital order - is also of immense importance for nanosized strongly correlated electron systems. 

In pure LaMnO$_3$, the orbital order superstructure develops due to degeneracy in 3d e$^1_g$ levels in Mn$^{3+}$O$_6$ octahedra which is lifted via Jahn-Teller effect resulting in a checkerboard-type cooperative ordering of the 3d$_{3x^2-r^2}$ and 3d$_{3y^2-r^2}$ orbitals within the ab-plane and their staggered in-phase stacking along the c-axis (so-called 'd'-type orbital order)\cite{Okamoto}. This orbital ordered state undergoes an order-disorder transition, reversibly, at a characteristic transition point T$_{JT}$\cite{Murakami}. This is accompanied by a structural transition \cite{Carvajal} as well as specific features \cite{Zhou} in different physical properties such as resistivity, thermoelectric power, Curie-Weiss paramagnetism, specific heat, thermal expansion coefficient etc. 

In this paper, we show that the thermodynamically stable long-range orbital order in stoichiometric LaMnO$_3$ evolves into a metastable order and finally a stable disorder as the particle size decreases from bulk ($\geq$1 $\mu$m) down to $\sim$10 nm. The orbital order-disorder transition temperature decreases significantly with the decrease in particle size. The transition becomes irreversible within a small window around a critical size d$_C$ $\sim$20 nm. This signifies that, within the window around d$_C$, the orbital ordered state is metastable. For even finer particles (e.g., $\sim$10 nm) beyond the window of metastable order, a stable orbital disordered phase evolves. The crystallographic phase too exhibits a systematic evolution as a function of particle size.  

\section{Experiments}
The nanoscale particles of phase pure LaMnO$_3$ system have been prepared via bio-template assisted synthesis. The carbon replicas of porous stems of pine wood were used as bio-templates. The walls of the pores (diameter $\sim$10-40 $\mu$m) appear to contain even finer porous channels with size range 5-10 nm (Fig. 1). The mixed metal nitrate aqueous solution is infiltrated within such templates under vacuum ($\sim$10$^{-2}$ torr). Following infiltration, the sample is first dried at 100$^o$C and then heat treated at 600$^o$-700$^o$C for 2-5h. The remnant after heat treatment is found to be single phase LaMnO$_3$. As the particle size reduces, the orthorhombic symmetry (space group Pbnm) of the bulk sample gives way to cubic (space group Pm$\bar{3}$m) with systematic decrease in orthorhombic distortion (Fig. 2). Along with the raw data, we also show in Fig. 2 the Rietveld refined patterns for selective cases. In Table-I, the room temperature lattice parameters (\textit {a, b, c}), the orthorhombic distortion \textit {s} [=(\textit{b}-\textit {a})/(\textit{b}+\textit{a})], the microstrain of the crystallites, and the reliability factors of the Rietveld refinement (by Fullprof ver 2.3) are listed. 

\begin{figure}[!h]
\centering
\includegraphics[scale=.6]{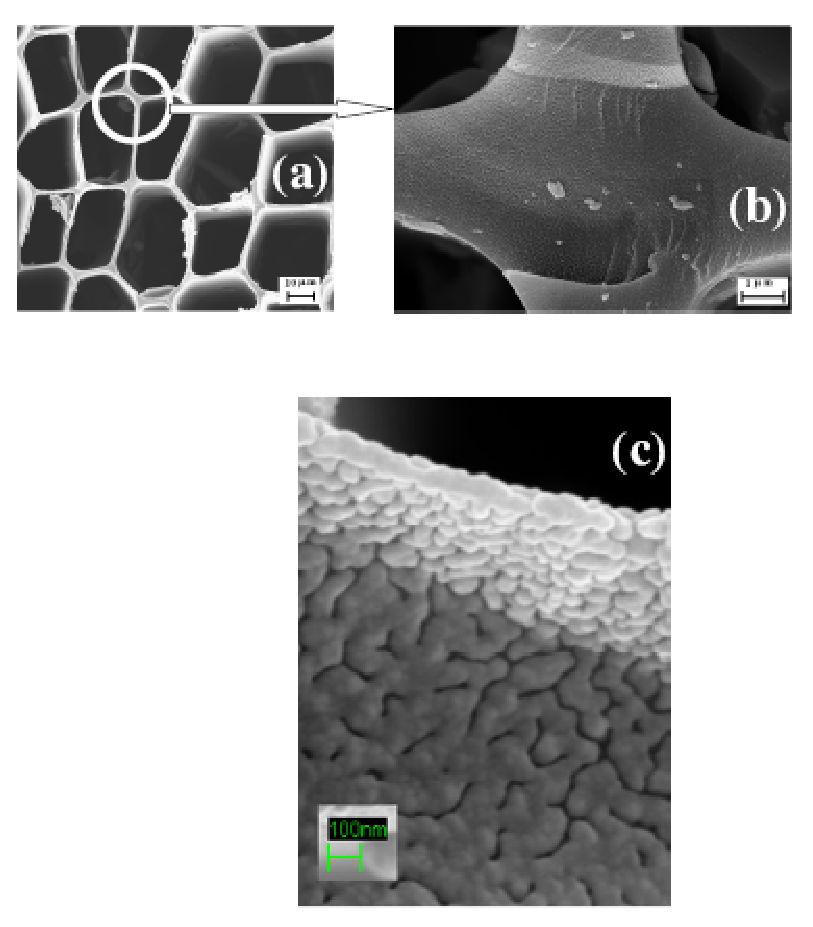}
\caption{(a) Field emission scanning electron micrograph (FESEM) of the porous carbon replica of bio-template (stem of pine wood), (b) blown up photograph of the wall of the bigger pore channels and (c) the finer pore structure (5-10 nm) of the wall.}
\end{figure}

\begin{table*}[!ht]
\caption{List of lattice parameters, orthorhombic distortion, and microstrain at room temperature. }

\begin{tabular}{p{0.8in}p{0.8in}p{0.8in}p{0.8in}p{0.8in}p{0.8in}p{0.8in}} \hline \hline
Samples & Lattice \newline parameters \newline (\AA) & Lattice \newline Volume \newline (\AA$^3$) &  Orthorhombic \newline distortion(s) (\%)  & Microstrain (\%) & $R_p$ & $\chi^2$ \\ \hline 
\\
LaMnO$_3$\newline -bulk & a = 5.532 \newline b = 5.748 \newline c = 7.683 & 245.305 & 1.915 & 0.008 & 20.1 & 1.2 \\  
\\
LaMnO$_3$ \newline - 30 nm & a = 5.529 \newline b = 5.678 \newline c = 7.708 & 242.039 & 1.329 & 0.047 & 21.68 & 1.518 \\ 
 \\ 
LaMnO $_3$\newline - 20 nm & a = 5.507 \newline b = 5.575 \newline c = 7.747 & 237.815 & 0.615 & 0.113 & 22.15 & 1.841 \\ 
 \\ 
LaMnO $_3$\newline -18 nm & a = 5.443 \newline b = 5.502 \newline c = 7.707 & 230.805 & 0.539 & 0.118 & - & - \\ 
\\ 
LaMnO $_3$\newline -12 nm & a = 3.878 &   58.095 & 0 & 0.126 & 14.53 & 1.18 \\  
\\ 
LaMnO $_3$\newline -8 nm & a = 3.862 & 57.587 & 0 & 0.216 & - & - \\ \hline \hline

\end{tabular}
\end{table*}

We prepared several samples with average particle size varying within $\sim$10-50 nm by varying the infiltration time, heat-treatment temperature and time. It is interesting to note that we observe presence of nanorods (dia $\sim$15-20 nm and above) as well as nanochains along with nearly spherical nanoparticles (Fig. 3). The particle size distribution and the average size for all the samples have been estimated from the field-effect scanning electron and transmission electron microscopy (Fig. 3) using the image analyzer software Image \textit{J}. In representative cases, the distribution patterns are shown (Fig. 3 insets). The orbital order-disorder transition data - transition point, electrical resistivity etc - are reported against the average particle size estimated from the distribution pattern for each sample. We carried out the electrical resistivity measurements directly on such nanoscale particles by preparing slurry of LaMnO$_3$ powder dispersed within an alumina sol. The slurry has been prepared by mixing colloidal suspension of alumina particles ($\sim$2\% by weight) with the LaMnO$_3$ particles ($\sim$98\% by weight). The slurry was then coated onto an alumina substrate. Gold electrodes and wires were already printed onto the substrate. After coating, the sample was heat-treated at $\sim$300$^o$C for 1h for curing. The adhesion of the coating onto the substrate and electrodes was found to be satisfactory. The resistivity ($\rho$) versus temperature (\textit{T}) measurements have been carried out on all such nanoscale samples and the data presented here are representative of the corresponding average particle size. The heat-treatment of the precursor powder as well as all the temperature and time-dependent electrical measurements were carried out in inert atmosphere (under flowing nitrogen) in order to avoid presence of non-stoichiometric oxygen which influences the orbital order-disorder transition significantly. The temperature-dependent x-ray diffraction measurements were carried out in vacuum ($\sim$10$^{-3}$ torr). 

\begin{figure}[!h]
\centering
\includegraphics[scale=.5]{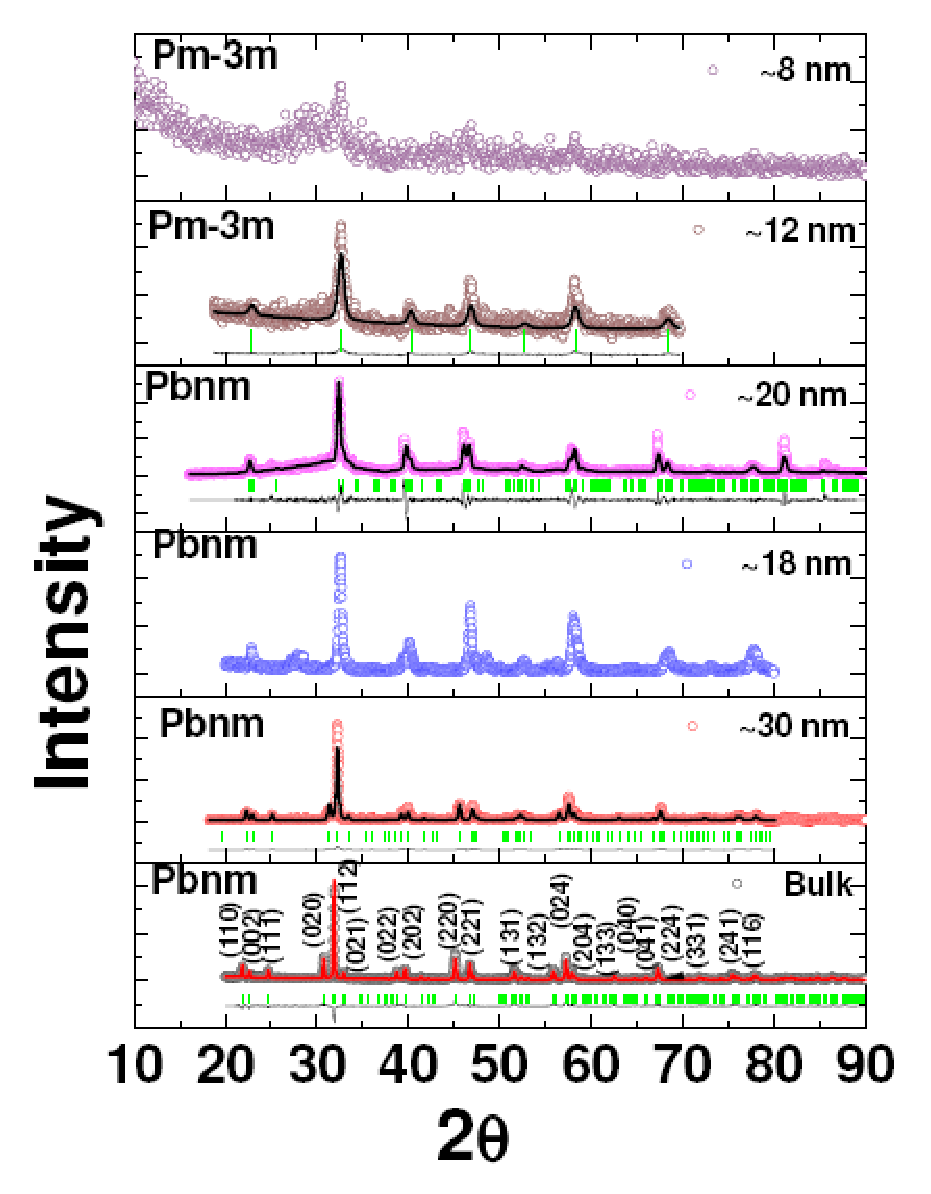}
\caption{(color online) Room temperature x-ray diffraction patterns (raw data as well as Rietveld refined ones) for the bulk and nanoscale LaMnO$_3$; the orthorhombic distortion decreases with the decrease in particle size giving way eventually to cubic symmetry in finer particles.}
\end{figure} 

The valence of Mn ions was found out in as-prepared nanoscale LaMnO$_3$ samples of different particle sizes from x-ray photoelectron spectroscopy measurement. The spectra were recorded at normal take off angle with pass energy 40 eV using Omicron Multiprobe Spectrometer (Omicron Nanotechnology, GmbH, UK) fitted with an EA125 hemispherical analyzer and a monochromatized Al K$\alpha$  source (1486.6 eV). 

\begin{figure}[!h]
\centering
\includegraphics[scale=.7]{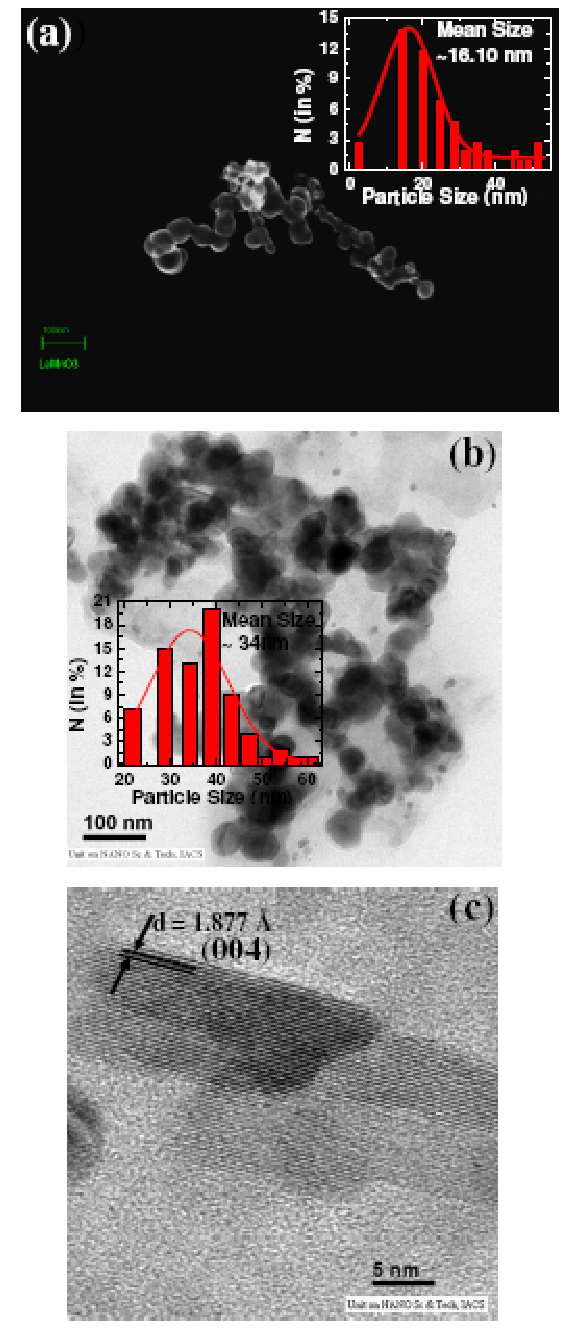}
\caption{(color online) (a) The FESEM image of a nearly $\sim$1 $\mu$m long nanochain of LaMnO$_3$ whose diameters are $\sim$20 nm at different places; (b) the transmission electron microscopic (TEM) image of nanoscale particles (diameter $\sim$34 nm) of LaMnO$_3$ (c) the high resolution transmission electron microscopy (HRTEM) image of the single crystalline nanorods of diameter 10-15 nm.}
\end{figure}

The spectra for Mn\textit{2p} and La\textit{3d} are shown in Fig. 4. The background of the spectra was subtracted by Shirley method \cite{Shirley}. The Mn\textit{2p} spectrum shows the presence of Mn\textit{2p}$_{3/2}$ and Mn\textit{2p}$_{1/2}$ peaks at 641.8 and 653.4 eV respectively with spin-orbit splitting of 11.6 eV corresponding to Mn$^{3+}$ in LaMnO$_3$. The values are in close agreement with those of Chainani et \textit{al}\cite{Chainani}. The FWHM of core level Mn\textit{2p}$_{3/2}$ spectra were found to be $\sim$3.3 eV for nanoscale LaMnO$_3$. 

\begin{figure}[!h]
\centering
\includegraphics[scale=.42]{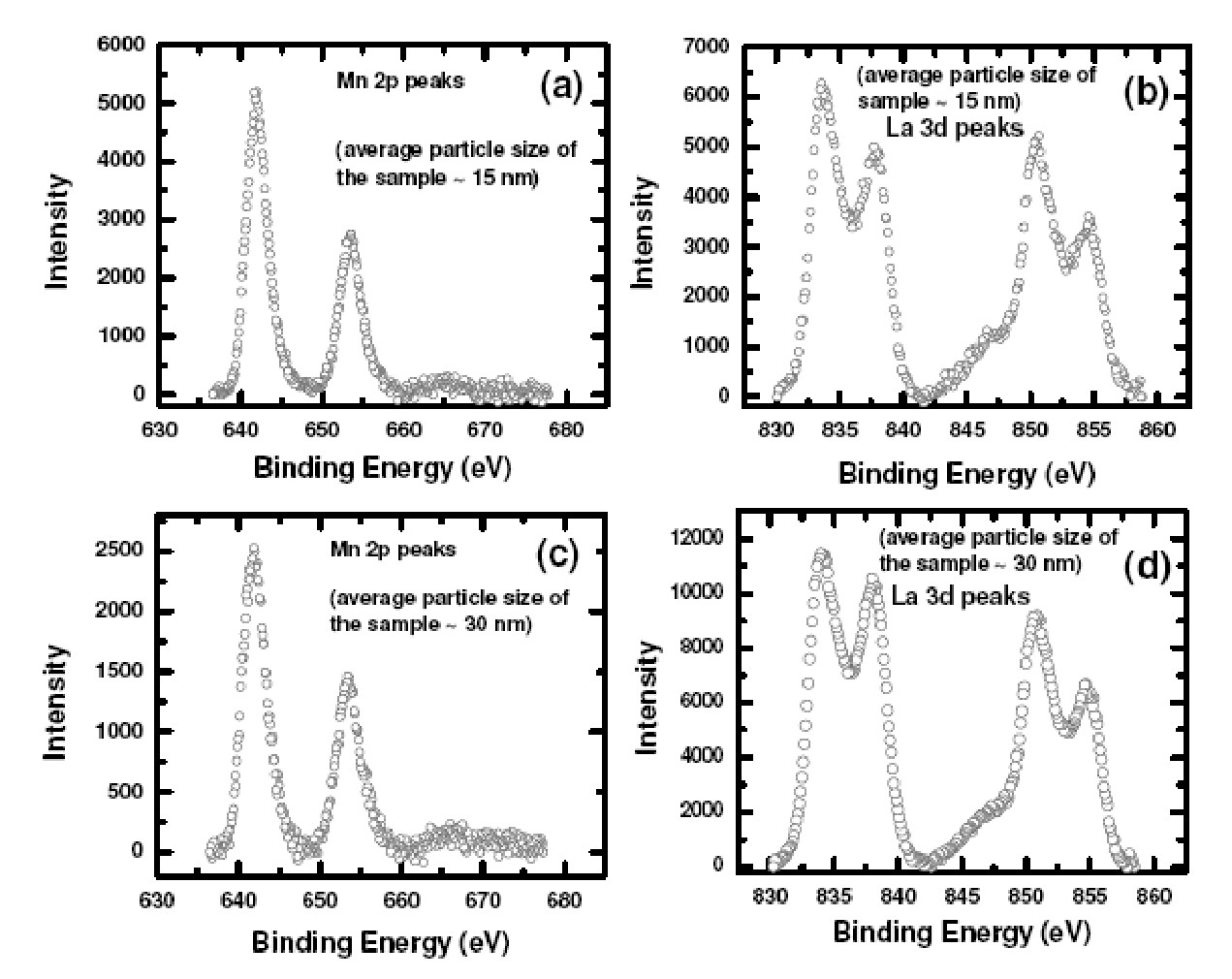}
\caption{(a) Mn 2p and (b) La 3d peaks and satellites in x-ray photoelectron spectroscopy for nanoscale LaMnO$_3$ (average particle size $\sim$15 nm); (c) Mn 2p and (d) La3d peaks and satellites for nanoscale LaMnO$_3$ (average particle size $\sim$30 nm).}
\end{figure}

In addition to the main Mn\textit{2p} peaks, a broad and low intensity satellite peak characteristic of LaMnO$_3$ was observed at 665.2 eV corresponding to Mn\textit{2p}$_{1/2}$ main peak at a separation of 11.8 eV. No additional peak corresponding to Mn$^{4+}$ has been observed. The La\textit{3d} spectra exhibit peaks and satellites corresponding to La\textit{3d}$_{5/2}$ and La\textit{3d}$_{3/2}$ as shown in Fig 4.  

\begin{figure*}[!ht]
\centering
\includegraphics[scale=.5]{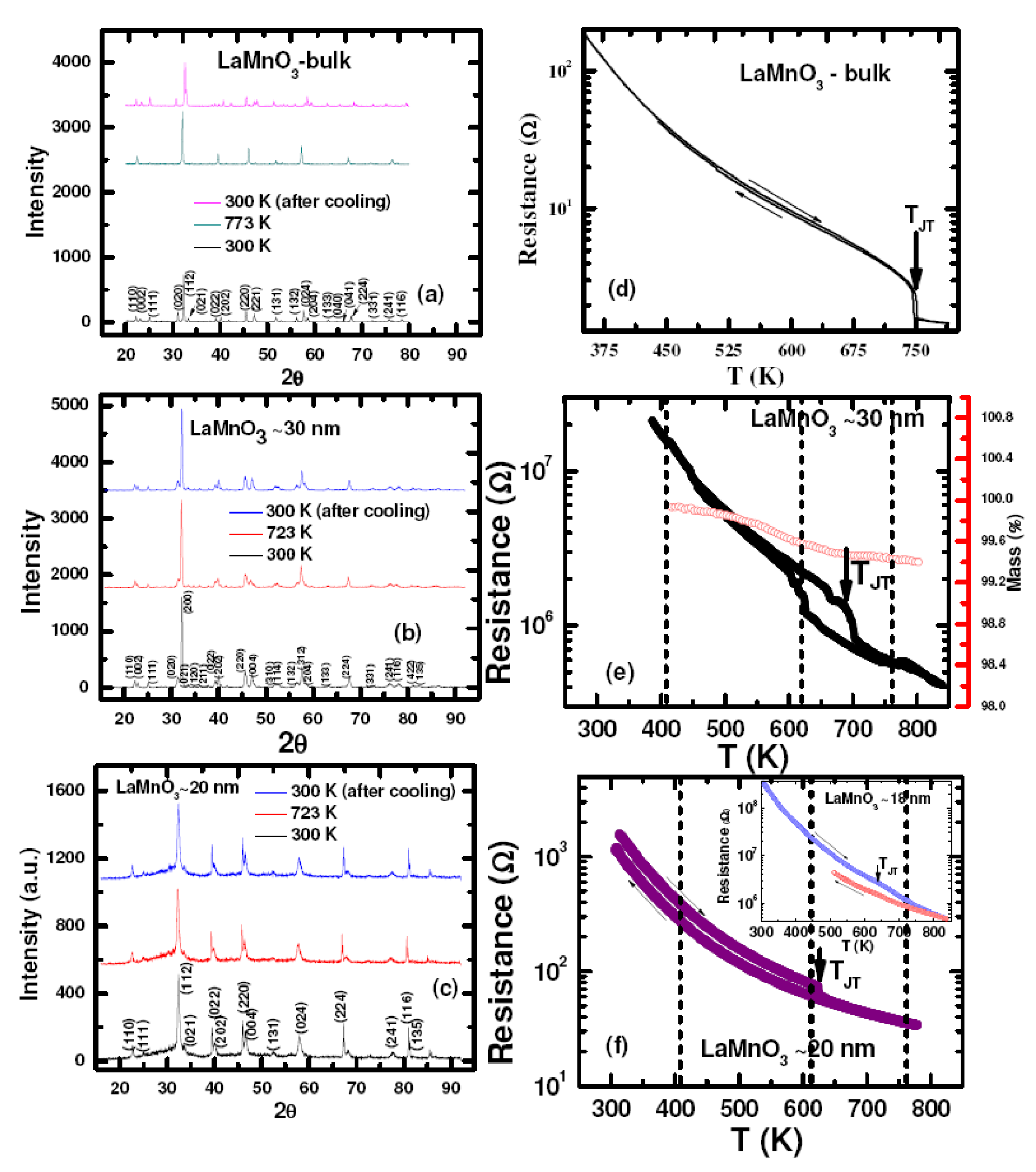}
\caption{(color online) The temperature-dependent x-ray diffraction patterns for (a) LaMnO$_3$-bulk, (b) $\sim$30 nm, and (c) $\sim$20 nm particles across T$_{JT}$s along with their corresponding $\rho$-$\textit{T}$ patterns in (e), (f), and (g), respectively. In (e), a representative thermogravimetry pattern is shown; the data points (open circles) and the corresponding right y-axis are of identical color. }
\end{figure*}

\section{Results and Discussion}
The characteristic features in the  $\rho$-\textit{T} patterns at respective T$_{JT}$s could be noticed. The T$_{JT}$s, thus found out, match closely with those identified from the peaks in the calorimetric studies \cite{Das} for the particles of average sizes above d$_C$ for which peaks in the calorimetry measurement appear. For the samples of size d$_C$ and less, no peak in the calorimetric study could be noticed. We have also tracked the transition by measuring the temperature-dependent x-ray diffraction patterns as orbital order-disorder transition in LaMnO$_3$ is associated with a structural phase transition. In Fig. 5, we show the  $\rho$-$\textit{T}$ as well as XRD patterns side-by-side to illustrate the compatibility between electrical and crystallographic data for a few samples which are important for mapping the entire phase diagram of orbital order-disorder transition versus particle size: (i) bulk, (ii) particles of average size $\sim$30 nm, and (iii) d$_C$ ($\sim$20 nm). While distinct signature of orbital order-disorder transition appears in the  $\rho$-\textit{T} plot at characteristic T$_{JT}$s, the crystallographic structure too depicts a clear feature - drop in the orthorhombic distortion s. The drop in $\textit{s}$ is expected for LaMnO$_3$ across the T$_{JT}$\cite{Carvajal}. The lattice parameters, volumes, orthorhombic distortions etc at different temperatures are listed in Table-II. Most interesting is the observation that while both the bulk and $\sim$30 nm particles exhibit reversible transition at respective T$_{JT}$s, the transition appears to be irreversible in the case of $\sim$20 nm particles. No signature of disorder-order transition could be noticed during the cooling cycle. Instead, the  $\rho$-$\textit{T}$ pattern observed at orbital disordered state beyond T$_{JT}$ appears to have been extended right down to $\sim$300 K. It might seem that perhaps disorder to order transition would take place at below 300 K. In that case, the transition would have been associated with a large hysteresis which should have given rise to a large endothermic peak in calorimetry. However, no such peak could be observed \cite{Das}. More importantly, a second run, under identical applied current, does not depict any signature of transition at all, at any temperature within 300-800 K. This cannot happen in the case of a first order transition. This is quite a new observation, as this has never been observed in the bulk system. The crystallographic structure too follows this pattern. The orthorhombic distortion drops above T$_{JT}$ but is not regained after the sample is cycled back to room temperature (Table-II). The structure of smaller distortion at above T$_{JT}$ appears to have been arrested. Therefore, the study of crystallographic data also shows that the orbital order-disorder transition becomes irreversible for the $\sim$20 nm particles. Additionally, we point out that, in contrast to the observations made in the case of $\sim$20 nm particles, $\sim$30 nm particles ($>$d$_C$) exhibit 'reversible' transition at T$_{JT}$ with a large hysteresis consistent with large latent heat of orbital order-disorder transition detected by calorimetry\cite{Das}. Therefore, the orbital order-disorder transition, as studied by resistivity, calorimetry, and temperature-dependent XRD for bulk as well as nanoscale samples, exhibit very interesting particle size dependent characteristics - particles of size larger than d$_C$ exhibit 'reversible' transition with finite latent heat and resistivity hysteresis whereas for particles of size d$_C$ and below, the transition becomes 'irreversible' with zero latent heat. We also mention here that even finer particles, of size $\sim$8-12 nm, exhibit cubic symmetry (space group Pm$\bar{3}$m) at room temperature (Fig. 2) and no signature of any transition within the 300-800 K regime (data not shown here). The T$_{JT}$ in this particle size regime could be shifted down to even below $\sim$300 K. This temperature region is presently not accessible to us for the XRD studies.

\begin{table*}[!ht]
\caption{List of lattice parameters and orthorhombic distortion at below and above T$_{JT}$. }
\begin{tabular}{p{0.6in}p{0.85in}p{0.85in}p{0.85in}p{0.85in}} \hline \hline
Samples & 300K & 723K & 773K & 300K (after cooling) \\ \hline 
LaMnO$_3$\newline -bulk & a = 5.532 \AA \newline b = 5.748 \AA \newline c = 7.683 \AA \newline V = 244.305 \AA$^3$ \newline s = 1.915  & - & a = 5.577 \AA \newline b = 5.577 \AA \newline c = 7.878 \AA \newline V = 245.029 \AA$^3$ \newline s = 0.0 & a = 5.531 \AA \newline b = 5.739 \AA \newline c = 7.683 \AA \newline V = 243.909 \AA$^3$ \newline s = 1.846 \\  
  
LaMnO$_3$ \newline - 30 nm & a = 5.529 \AA \newline b = 5.678 \AA \newline c = 7.708 \AA \newline V = 242.039 \AA$^3$ \newline s = 1.329  &  a = 5.543 \AA \newline b = 5.551 \AA \newline c = 7.873 \AA \newline V = 243.665 \AA$^3$ \newline s = 0.0721 & - & a = 5.529 \AA \newline b = 5.672 \AA \newline c = 7.704 \AA \newline V = 241.641 \AA$^3$ \newline s = 1.277  \\ 
 
LaMnO$_3$\newline - 20 nm & a = 5.507 \AA \newline b = 5.575 \AA \newline c = 7.747 \AA \newline V = 237.815 \AA$^3$ \newline s = 0.615 & a = 5.514 \AA \newline b = 5.528 \AA \newline c = 7.805 \AA \newline V = 237.915 \AA$^3$ \newline s = 0.127 & - & a = 5.516 \AA \newline b = 5.532 \AA \newline c = 7.798  \AA \newline V = 237.952 \AA$^3$ \newline s = 0.145 \\  \hline \hline

\end{tabular}
\end{table*}

In order to gather more information about the stability of the orbital ordered state in samples of size around d$_C$ ($\sim$20 nm), we measured the relaxation characteristics - i.e., time dependence of the resistivity - at below and above T$_{JT}$ following two different protocols: (i) heating a fresh sample to the respective points and then initiating the relaxation measurement after stabilizing the temperature for nearly $\sim$30s and (ii) heating a fresh sample beyond T$_{JT}$ up to $\sim$800 K and then cooling down to the respective points. The heating and cooling rates were kept fixed at $\sim$1.5$^o$C/min, which were used for simple  $\rho$-$\textit{T}$ measurements as well. Identical heating/cooling rate and stabilization time for both the set of measurements ensures identical initial condition at any particular temperature. The temperatures chosen for the relaxation studies are: (i) $\sim$410 K (i.e., well below T$_{JT}$), (ii) $\sim$618 K (i.e., near the T$_{JT}$) and (iii) $\sim$765 K (i.e., well above T$_{JT}$). We repeated the relaxation studies both in the particles of size above d$_C$ and in a single crystal by choosing the temperatures appropriately. The relaxation was measured by reaching and stabilizing the temperature first and then by triggering the current flow and voltage recording simultaneously. The data were recorded at a time interval of $\sim$43 ms over a time span of $\sim$3000s. The relaxation characteristics for $\sim$20 nm particles are shown in Figs. 6a. Quite evident in Fig. 6a is the fact that as the temperature is increased from well below T$_{JT}$ to close to T$_{JT}$, the decay rate of the resistivity increases appreciably. However, well above T$_{JT}$ no decay in resistivity could be noticed. Likewise, no decay in resistivity could be noticed when a fresh sample is cooled down to the lower temperatures from above T$_{JT}$ and relaxation measurement is done at those points. This observation further supports the conjecture that the orbital ordered state in samples of size around d$_{C}$ is metastable. The decay rate enhances near T$_{JT}$ because of enhanced fluctuations arising from onset of transition dynamics in a metastable phase. The disordered state is quite stable and depicts no appreciable decay in resistivity. The single crystal and the particles of size greater than d$_C$ (i.e., say, $\sim$30 nm) depict stable ordered and disordered states and therefore no such time decay of resistivity (Figs. 6b,c). 

\begin{figure}[!ht]
\centering
\includegraphics[scale=.5]{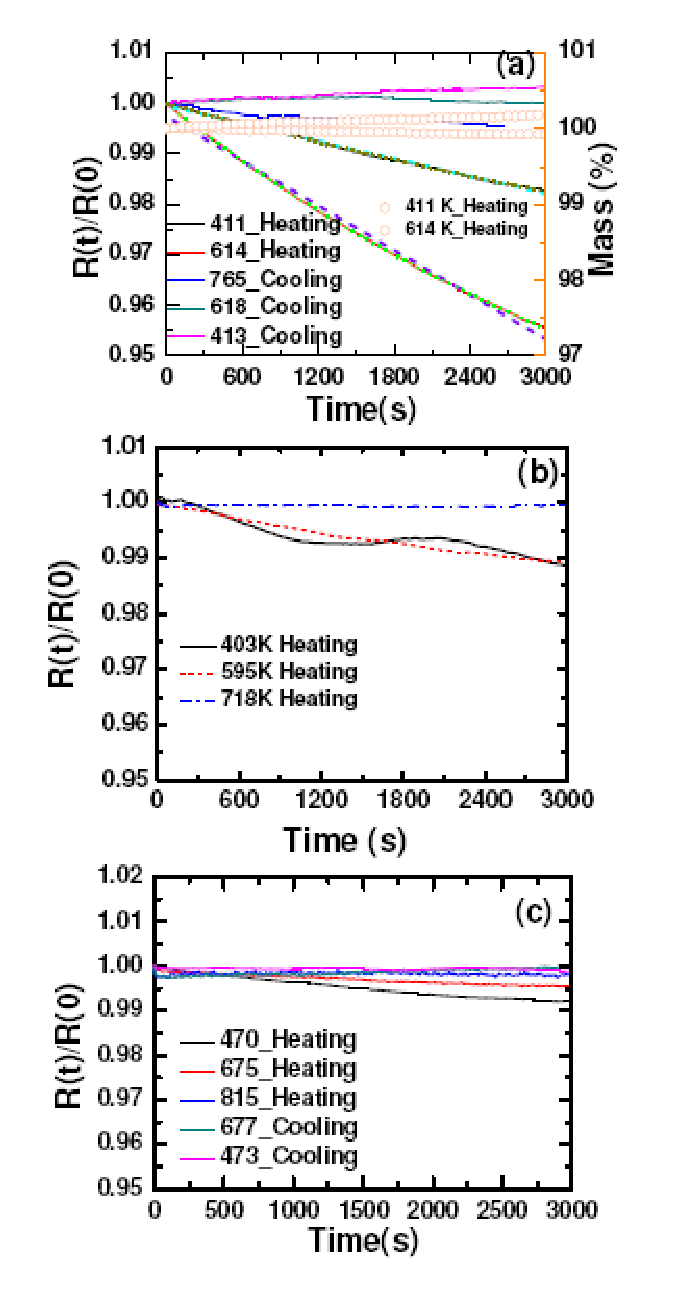}
\caption{(color online) The time decay characteristics of the resistivity for (a) $\sim$20 nm particles, (b) $\sim$30 nm particles, and (c) single crystal LaMnO$_3$ at different temperatures in the orbital ordered and disordered phases reached via two different protocols; the temperatures at which the time-dependent resistance was measured are marked in Figs. 5(e),(f) by dashed lines; the weight loss/gain for the nanoscale samples can be noted from the representative thermogravimetry data shown for $\sim$20 nm particles as a function of time in (a); the data points (open circles) and the corresponding right y-axis can be identified by appropriate color matching. }
\end{figure}

\begin{figure}[!h]
\centering
\includegraphics[scale=.5]{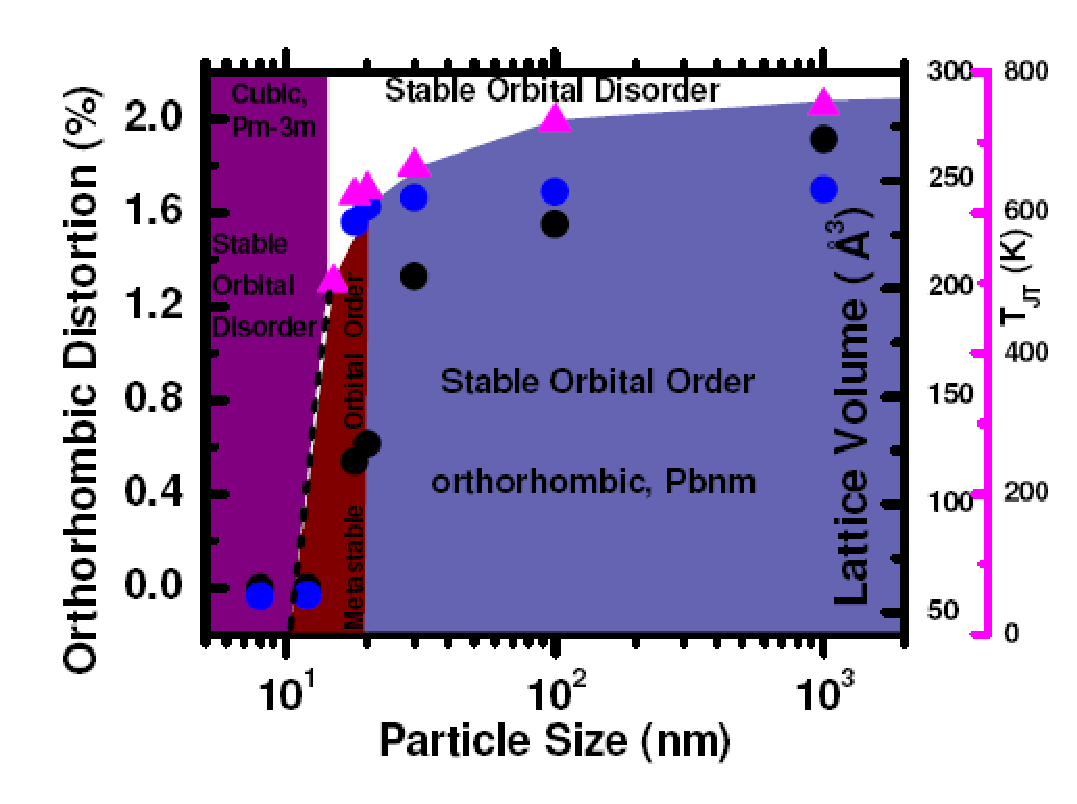}
\caption{(color online) A phase diagram of different orbital phases in LaMnO3 as a function of particle size - across a wide range from bulk ($>$1$\mu$m) down to nanosize ($\sim$1 nm) - has been constructed using orthorhombic distortion (black circles), lattice volume (blue circles), and T$_{JT}$ (up-triangles) data for different particle sizes; while the boundary between metastable and stable orbital order has been identified in this work, the boundary between metastable orbital ordered and stable orbital disordered phases at even smaller particle size range ($<$10 nm) is tentative and is yet to be firmly established. }
\end{figure}

The decay pattern of resistivity, in particles of size d$_C$ at $\sim$411 K, appears to be approximately linear with time while the pattern at $\sim$618 K tends more toward an exponential one. We have calculated the activation energy ($\textit{E}$) of decay by fitting the data for $\sim$411 K with R(t)/R(0) = c - (k$_B$T/E).(t/$\tau$ ) and the ones for $\sim$618 K with R(t)/R(0) = c - (k$_B$T/E).exp(t/$\tau$ ) where $\textit{c}$ is a constant and $\tau$ is a characteristic relaxation time scale. The quality of fit of the experimental data with linear (dashed line) and exponential (dotted line) decay equations can be seen in the plot of Fig. 6a. The activation energy is found to decrease from $\sim$36 meV to $\sim$6 meV over a temperature range 411-618 K. The 'linear in time relaxation pattern' was observed\cite{Corruccini} for recovery of magnetization from saturation in the superfluid $^3$He. No theoretical prediction about 'orbital domain' fluctuations and consequent relaxation in resistivity below a critical size for LaMnO$_3$ or other orbital ordered compounds exists now. The experimental results being presented here will be important in formulating a theory of relaxation of orbital ordered state in nanoscale LaMnO$_3$. It is important to mention, in this context, that there is a crucial difference between fluctuating orbital ordered states observed in nanoscale ($\sim$20 nm) LaMnO$_3$ and fluctuating superparamagnetic/superparaelectric states in nanosized ferromagnetic/ferroelectric systems above blocking temperature (T$_B$). Whereas in the latter case fluctuation dynamics is faster above T$_B$ and, therefore, temperature scanning of magnetization/polarization does not reveal any transition point (Curie point) at all, in the former case, the fluctuation appears to be much slower to reveal a transition temperature, at least, in the very first thermal cycling of a fresh sample. Of course, no signature of orbital order-disorder transition could be observed in any subsequent thermal cycle or in measurement carried out over longer time scale (3000s). The fluctuation can increase or, in other words, the time scale of fluctuation can decrease in even finer sized samples.

In order to find out whether oxygen intake or release from the nanosized particles - during the resistivity measurement experiments as a function of temperature and time - is influencing the resistivity data or not in the present case, we carried out the thermogravimetric measurement of the samples under identical conditions. The temperature and time dependent data are plotted in Figs. 5e and 6a, respectively. The weight loss or gain in both temperature as well as time dependent experiment is found to be negligible ($\sim$0.4$\%$). This result together with the x-ray photoelectron spectroscopy data (Fig. 4) on as-prepared powder shows that the LaMnO$_3$ particles retain proper oxygen stoichiometry throughout the course of the work. On the contrary, the non-stoichiometric LaMnO$_3$ exhibits more than 2-4$\%$ oxygen gain or loss in thermogravimetric measurement from which one can estimate the extent of non-stoichiometry in oxygen\cite{Andersen}. Therefore, it is clear that the time-decay in resistivity observed in the nanosized particles (size d$_C$ $\sim$20 nm) is a reflection of intrinsic metastability of the orbital ordered state. The metastable orbital ordered state decays and tends toward the stable orbital disordered state with 2-4$\%$ decrease in resistance within a time span of $\sim$3000s.

It has been shown earlier by others\cite{Sarkar}$^,$\cite{Raychaudhuri} that the charge-ordered systems undergo a melting-type transition in nanoscale because of enhanced pressure effect. In fact, the phenomenon of melting of the charge ordered state under pressure, associated with insulator to metal and antiferromagnetic to ferromagnetic transitions, appears to have been replicated well in the nanoscale system. In the case of LaMnO$_3$ with orbital order, it seems that although pressure does increase in the nanoscale sample because of enhanced surface area to volume ratio as the particle size is decreased from $\sim$100 nm down to $\sim$1 nm, melting from a stable ordered state to a stable disordered state takes place via an intermediate metastable state where the pressure is not yet sufficient for melting. Opening up of this small window of metastable orbital order at an intermediate particle size range is quite interesting for this orbital ordered compound and is in contrast to the observations made in charge-ordered compounds. The frequency of fluctuation of the orbital order within the metastable zone of the phase diagram increases with the decrease in particle size. We did not, of course, observe any metal-insulator transition or ferromagnetism in nanoscale LaMnO$_3$ at sizes d$_C$ or below as reported for charge ordered compounds by others\cite{Sarkar}$^,$\cite{Raychaudhuri}. Even, much finer particles (average size $\sim$8-12 nm $\ll$ d$_C$), which do not depict any signature of transition within 300-800 K, and, therefore, could be in the orbital disordered state within this temperature regime, did not depict metal-insulator transition or ferromagnetism. One of the reasons could be that the orbital disordered state in bulk LaMnO$_3$ too, does not depict good metallic or ferromagnetic behavior \cite{Zhou}. Even the pressure driven melting of Jahn-Teller order also does not yield a good metallic or ferromagnetic phase \cite{Loa}$^,$\cite{Zhou-2}. Instead, the magnetic phase could possibly be more akin to what has been observed in the case of LaMn$_{0.5}$Ga$_{0.5}$O$_3$ system \cite{Zhou-3} where breakdown of long-range orbital order has shown to have given rise to spin glass phase. A detailed study of the magnetic structure for $\sim$10-20 nm particles of LaMnO$_3$, where long-range orbital order appears to have become metastable or have broken down, will be carried out in near future.

The underlying crystallographic structure depicts a systematic evolution with the decrease in particle size. The lattice volume and orthorhombic distortion decrease and the crystal symmetry transforms from orthorhombic to cubic. This is consistent with the picture of breakdown of long-range orbital order and stabilization of orbital disordered phase in $\sim$10 nm particles as orbital disordered phase stabilizes in cubic system with smaller lattice volume \cite{Chatterji}. Using the lattice volume, orthorhombic distortion, and T$_{JT}$ values, we construct a phase diagram (Fig. 7) which shows how a thermodynamically stable long-range orbital ordered phase evolves into a stable disordered phase via an intermediate metastable orbital ordered phase with particle size. Such a map of evolution of orbital phases with particle size in nanoscale is $\textit{the central result of this paper}$. 

\section{Summary}
In summary, we found that the orbital ordered state in pure LaMnO$_3$ becomes metastable for particles of size around d$_C$ ($\sim$20 nm) and finally transforms into a stable disordered state for much finer particles (e.g., $\sim$10 nm). The orbital order-disorder transition switches from 'reversible' to 'irreversible' at d$_C$ along with sizable decay in resistivity with time in the ordered state. In contrast to the observation made in the case of charge ordered compounds, we did not observe an insulator-metal transition and antiferromagnetic-ferromagnetic transition for a size around d$_C$. The crystal symmetry, lattice volume, and orthorhombic distortion depict a systematic variation with particle size. The lattice volume collapses and the orthorhombic symmetry gives way to cubic as the particle size reduces from bulk down to $\sim$10 nm signifying systematic evolution from long-range orbitally ordered orthorhombic phase to orbitally disordered cubic phase.

\textbf{ACKNOWLEDGMENTS.} 
The authors thank P. Choudhury and J. Ghosh for helpful discussion and P. Mandal for the single crystal of LaMnO$_3$. One of the authors (PM) acknowledges financial support from CSIR while another author (OPC) thanks DST for a sponsored project.


\begin{thebibliography} {99} 

\bibitem{Buffat} Ph. Buffat and J.-P. Borel, Phys. Rev. A $\textbf{13}$, 2287 (1976). 
\bibitem{Dippel} M. Dippel, A. Maier, V. Gimple, H. Wider, W.E. Evenson, R.L. Rasera, and G. Schatz, Phys. Rev. Lett. $\textbf{87}$, 095505 (2001).
\bibitem{Zhang} M. Zhang, M.Y.  Efremov, F. Schiettekatte, E.A. Olson, A.T.  Kwan, S.L. Lai, T. Wisleder, J.E. Green, and L.H. Allen, Phys. Rev. B $\textbf{62}$, 10548 (2000).
\bibitem{Schmidt} M. Schmidt, R. Kusche, B. von Issendorf, and H. Haberland, Nature (London), $\textbf{393}$, 238 (1998).
\bibitem{Tokura} See, for example, Y. Tokura and N. Nagaosa, Science $\textbf{280}$, 462 (2000).
\bibitem{Okamoto} S. Okamoto, S. Ishihara, and S. Maekawa, Phys. Rev. B $\textbf{65}$, 144403 (2002); E. Pavarini and E. Koch, Phys. Rev. Lett. $\textbf{104}$, 086402 (2010). 
\bibitem{Murakami} Y.  Murakami, J.P. Hill, D. Gibbs, M.  Blume, I. Koyama, M. Tanaka, H. Kwata, T. Arima, Y. Tokura, K. Hirota, and Y. Endoh, Phys. Rev. Lett. $\textbf{81}$, 582 (1998).
\bibitem{Carvajal} J.R. Carvajal, M.  Hennion, F. Moussa, A.H. Moudden, L. Pinsard, and A. Revcolevschi, Phys. Rev. B $\textbf{57}$, R3189 (1998).
\bibitem{Zhou} See, for example, J.-S.  Zhou and J.B. Goodenough, Phys. Rev. B $\textbf{60}$, R15002 (1999); see also, D. Bhattacharya and H.S. Maiti, Phys. Rev. B $\textbf{66}$, 132413 (2002).
\bibitem{Shirley} D.A. Shirley, Phys. Rev. B $\textbf{5}$, 4709 (1972). \bibitem{Chainani} A. Chainani, M. Mathew, and D.D. Sarma, Phys. Rev. B $\textbf{47}$, 15397 (1993).
\bibitem{Das} N. Das, P. Mondal, and D. Bhattacharya, Phys. Rev. B $\textbf{74}$, 014410 (2006).
\bibitem{Corruccini} See, for example, L.R. Corruccini and D.D. Osheroff, Phys. Rev. Lett. $\textbf{34}$, 564 (1975).
\bibitem{Andersen} I.G.K. Andersen, E.K. Andersen, P. Norby, and E. Skou, J. Solid State Chem. $\textbf{113}$, 320 (1994).
\bibitem{Sarkar} T. Sarkar, B. Ghosh, A.K. Raychaudhuri, and T. Chatterji, Phys. Rev. B $\textbf{77}$, 235112 (2008).
\bibitem{Raychaudhuri} T. Sarkar, A.K. Raychaudhuri, and T. Chatterji, Appl. Phys. Lett. $\textbf{92}$, 123104 (2008).
\bibitem{Loa} I. Loa, P. Adler, A. Grzechnik, K. Syassen, U. Schwarz, M. Hanfland, G.Kh. Rozenberg, P. Gorodetsky, and M.P. Pasternak, Phys. Rev. Lett. $\textbf{87}$, 125501 (2001). 
\bibitem{Zhou-2} J.-S. Zhou and J.B. Goodenough, Phys. Rev. Lett. $\textbf{89}$, 087201 (2002). 
\bibitem{Zhou-3} J.-S. Zhou and J.B. Goodenough, Phys. Rev. B $\textbf{68}$, 144406 (2003). 
\bibitem{Chatterji} T. Chatterji, F. Fauth, B. Ouladdiaf, P. Mandal, and B. Ghosh, Phys. Rev. B $\textbf{68}$, 052406 (2003). 

\end{thebibliography}
\end{document}